\documentclass[12pt]{iopart}
\usepackage{iopams,setstack}
\newcommand{\vect}[1]{\bi{#1}}
\begin{document}
\title[Hamiltonian for semiconductor nanostructures in a magnetic
field]{First-principles effective-mass Hamiltonian for semiconductor
nanostructures in a magnetic field}
\author{Bradley A Foreman}
\address{Department of Physics, Hong Kong University of Science and
Technology, Clear~Water~Bay, Kowloon, Hong~Kong, China}
\ead{phbaf@ust.hk}
\begin{abstract}
A multi-band effective-mass Hamiltonian is derived for lattice-matched
semiconductor nanostructures in a slowly varying external magnetic
field.  The theory is derived from the first-principles magnetic-field
coupling Hamiltonian of Pickard and Mauri, which is applicable to
nonlocal norm-conserving pseudopotentials in the local density
approximation to density functional theory.  The pseudopotential of
the nanostructure is treated as a perturbation of a bulk reference
crystal, with linear and quadratic response terms included in
$\vect{k} \cdot \vect{p}$ perturbation theory.  The resulting
Hamiltonian contains several interface terms that have not been
included in previous work on nanostructures in a magnetic field.  The
derivation provides the first direct analytical expressions showing
how the coupling of the nonlocal potential to the magnetic field
influences the effective magnetic dipole moment of the electron.
\end{abstract}
\submitto{\JPCM}
\maketitle

\section{Introduction}

\label{sec:intro}

Recent experiments on the spin Hall effect \cite{Kato04,WuKaSiJu05}
and the spin galvanic and circular photogalvanic effects
\cite{Gan03c,Gan04a} have stimulated renewed interest in the behavior
of electrons and holes in semiconductor nanostructures in an external
magnetic field.  Such experiments are often modeled using an
effective-mass or envelope-function Hamiltonian \cite{Bast88,IvPi97}
that may include a generalized Rashba coupling \cite{Wink03} to
account for spin-splitting effects.  The magnetic field is included
via the minimal coupling substitution $\vect{k} \rightarrow \vect{k} +
\vect{A} / c$, which is well justified for bulk semiconductors
\cite{LuKo55,Lutt56,BirPik74}.

However, Mlinar \etal \cite{Mlinar05} have recently demonstrated that
the results of such calculations can depend strongly on the particular
choice of envelope-function Hamiltonian.  The standard ``symmetrized''
Hamiltonian \cite{EpScCo87} applied to GaAs/AlAs and InAs/GaAs quantum
dots in a strong magnetic field yielded energy eigenstates with
arguably unphysical properties \cite{Mlinar05}---in contrast with the
physically reasonable predictions \cite{Mlinar05} obtained from an
asymmetric Hamiltonian \cite{Fore93,Fore97}.  This may be viewed as a
modern confirmation of Luttinger's original work \cite{Lutt56}
demonstrating the importance of including antisymmetric terms in the
Hamiltonian.

Nevertheless, it is not clear that the model used by Mlinar \etal
\cite{Mlinar05} provides a correct description of the heterojunction
regions.  The antisymmetric part of the valence-band Hamiltonian
derived in reference \cite{Fore93} is in fact just the valence-band
Rashba coupling \cite{Wink03} that is of interest for spintronics
applications \cite{BernZhang04a}.  However, a recent first-principles
envelope-function theory for lattice-matched heterostructures
\cite{Fore05b} has established the existence of several other
interface terms of the same order of magnitude as the Rashba term.
These terms are not included in standard effective-mass models for
nanostructures, although some authors \cite{Rod03} have included
phenomenological surface terms in their study of the magnetic field
problem.

The purpose of this paper is to establish a rigorous theoretical
foundation for the effective-mass theory of semiconductor
nanostructures in a magnetic field.  This is done by extending the
first-principles multi-band effective-mass Hamiltonian of reference
\cite{Fore05b} to include the effects of a slowly varying static
external magnetic field.  In this theory \cite{Fore05b}, the
nanostructure pseudopotential is treated as a perturbation of a bulk
reference crystal (e.g.,
$\mathrm{Al}_{0.5}\mathrm{Ga}_{0.5}\mathrm{As}$ for a GaAs/AlAs
quantum dot), with the self-consistent linear and quadratic response
terms included in the framework of the $\vect{k} \cdot \vect{p}$
perturbation theory of Leibler \cite{Leib77}.

Until recently, such a first-principles analysis would not have been
possible, due to the lack of a generally accepted Hamiltonian to
describe the coupling between the magnetic field and the nonlocal part
of the potential energy.  Such nonlocal terms arise for two reasons.
At the fundamental level, the self-energy operator (which accounts for
electron-electron Coulomb interactions) in Dyson's equation is
nonlocal.  At the practical level, the pseudopotentials used in most
first-principles calculations (which account for the influence of core
electrons) are also nonlocal.

Pickard and Mauri \cite{PiMa03} have solved the second problem by
deriving a coupling for norm-conserving pseudopotentials in the local
density approximation (LDA) \cite{Payne92} that reproduces the results
of all-electron LDA calculations.  Ismail-Beigi \etal
\cite{IsChLouie01} have proposed a more general coupling for arbitrary
nonlocal potentials, but since this coupling, in the special case of
norm-conserving pseudopotentials, does not agree with all-electron
calculations \cite{PiMa03}, its validity in the general case is
questionable.

The present work is therefore limited to the case of norm-conserving
pseudopotentials within LDA.  Although LDA calculations do not
accurately predict energy gaps in semiconductors, this at least
provides a working model that is self-consistent and accounts for the
true atomic structure of a heterointerface.  The results obtained here
should be representative of the general features obtained in a more
complete theory that includes the nonlocal quasiparticle self-energy.

The results derived in this paper confirm the existence of interface
terms in the Hamiltonian that were not included in the calculations of
Mlinar \etal \cite{Mlinar05}.  An explicit numerical calculation of
these terms is in development and will be reported elsewhere.
However, the present analytical results show that these terms are of
the same order of magnitude as the valence-band Rashba terms studied
in reference \cite{Mlinar05}, which suggests that they may have a
comparable influence on the energy eigenvalues and eigenstates in an
external magnetic field.

In addition, an analytical demonstration of the difference between the
coupling Hamiltonians of Pickard and Mauri \cite{PiMa03} and
Ismail-Beigi \etal \cite{IsChLouie01} is established for the first
time.  In particular, the nonlocal pseudopotential contributes to the
effective magnetic dipole moment of the electron in the first case,
while in the second case it does not.

The paper begins in section \ref{sec:hamiltonian} with the derivation
of exact expressions for the magnetic-field coupling Hamiltonian in
coordinate and momentum space.  Approximations suitable for slowly
varying envelope functions are developed in section \ref{sec:approx},
based on a power series expansion of the nonlocal pseudopotential in
momentum space.  These results are used to construct an
envelope-function Hamiltonian in section \ref{sec:ef}.  The
contributions derived from $\vect{k} \cdot \vect{p}$ perturbation
theory are added in section \ref{sec:other}, where the complete
expression for the effective-mass Hamiltonian is given.  The results
are discussed in the concluding section \ref{sec:conclusions}.
Throughout the paper, Hartree atomic units ($\hbar = m = e = 1$) are
used.

\section{General form of the coupling Hamiltonian}

\label{sec:hamiltonian}

The magnetic coupling Hamiltonian derived by Pickard and Mauri
\cite{PiMa03} is applicable to nonlocal potentials formed by a
superposition of localized nonoverlapping norm-conserving
pseudopotentials; i.e.,
\begin{equation}
   V^{{\rm nl}} (\vect{x}, \vect{x}') = \sum_{\vect{R}}
   V_{\vect{R}}^{{\rm nl}} (\vect{x}, \vect{x}') ,
\end{equation}
where $V_{\vect{R}}^{{\rm nl}} (\vect{x}, \vect{x}')$ is the nonlocal
part of the pseudopotential for an ion at position $\vect{R}$.  In
this case, the potential in the presence of a slowly varying magnetic
field $\vect{B} (\vect{x}) = \nabla \times \vect{A} (\vect{x})$ is
given by \cite{PiMa03}
\begin{equation}
   \tilde{V}^{{\rm nl}} (\vect{x}, \vect{x}') = \sum_{\vect{R}}
   V_{\vect{R}}^{{\rm nl}} (\vect{x}, \vect{x}') \exp \Biggl(
   \frac{\rmi}{c} \int_{\vect{x} \rightarrow \vect{R} \rightarrow
   \vect{x}'} \vect{A} ( \vect{x}'' ) \cdot \rmd \vect{x}'' \Biggr) ,
   \label{eq:VA}
\end{equation}
where $\vect{x} \rightarrow \vect{R}$ denotes an integral over the
straight-line path from $\vect{x}$ to $\vect{R}$.  If $\vect{A}
(\vect{x})$ is chosen to be periodic \cite{BirPik74,PiMa03,MauLou96},
it can be expanded in a Fourier series, which permits the line
integral to be evaluated as
\begin{equation}
   \int_{\vect{R} \rightarrow \vect{x}'} \vect{A} \cdot \rmd
   \vect{x}'' = (\vect{x}' - \vect{R}) \cdot \sum_{\vect{k}} \vect{A}
   (\vect{k}) \rme^{\rmi \vect{k} \cdot \vect{R}} f [\rmi \vect{k}
   \cdot (\vect{x}' - \vect{R}) ] ,
\end{equation}
in which
\begin{equation}
   f(x) = \frac{\rme^x - 1}{x} = 1 + \frac{x}{2!} + \frac{x^2}{3!} +
   \cdots .  \label{eq:fx}
\end{equation}
Since the coupling (\ref{eq:VA}) is valid only for slowly varying
$\vect{B} (\vect{x})$ \cite{PiMa03}, it is assumed that $\vect{A}
(\vect{k})$ is significant only for small $k$ (i.e., $ka \ll 1$, where
$a$ is the lattice constant).  The power series in (\ref{eq:fx})
therefore converges rapidly, because the range of the nonlocal
potential (i.e., the core radius) is less than $a$.  From this power
series one obtains
\begin{eqnarray}
   \fl \int_{\vect{R} \rightarrow \vect{x}'} \vect{A} \cdot \rmd
   \vect{x}'' = (\vect{x}' - \vect{R}) \cdot \vect{A} (\vect{R}) +
   \frac{1}{2!}  (x_{\lambda}' - R_{\lambda}) (x_{\mu}' - R_{\mu})
   \frac{\partial A_{\lambda}}{\partial R_{\mu}} \nonumber \\ {} +
   \frac{1}{3!}  (x_{\lambda}' - R_{\lambda}) (x_{\mu}' - R_{\mu})
   (x_{\nu}' - R_{\nu}) \frac{\partial^2 A_{\lambda}}{\partial R_{\mu}
   \partial R_{\nu}} + \cdots , \label{eq:path1}
\end{eqnarray}
in which a sum over repeated Cartesian coordinate indices $\lambda$,
$\mu$, $\nu$ is assumed, and $\partial A_{\lambda} / \partial R_{\mu}
\equiv (\partial A_{\lambda} / \partial x_{\mu})|_{\vect{x} =
\vect{R}}$.  Therefore
\begin{eqnarray}
   \fl \int_{\vect{x} \rightarrow \vect{R} \rightarrow \vect{x}'}
   \vect{A} \cdot \rmd \vect{x}'' = (\vect{x}' - \vect{x}) \cdot
   \vect{A} (\vect{R}) \nonumber \\ {} + \frac{1}{2!} [ (x_{\lambda}'
   - R_{\lambda}) (x_{\mu}' - R_{\mu}) - (x_{\lambda} - R_{\lambda})
   (x_{\mu} - R_{\mu}) ] \frac{\partial A_{\lambda}}{\partial R_{\mu}}
   + \cdots .  \label{eq:path2}
\end{eqnarray}
Note that equations (\ref{eq:path1}) and (\ref{eq:path2}) depend only
on the symmetric part of the matrix $\partial A_{\lambda} / \partial
R_{\mu}$; i.e., they contain no terms proportional to $\vect{B}$ or
its derivatives.  As shown in the Appendix, this is generally not true
for other coupling Hamiltonians.

For the potential energy of a lattice-matched heterostructure, it is
convenient to use a more explicit notation:
\begin{equation}
   V(\vect{x}, \vect{x}') = \sum_{\alpha, \vect{R}}
   f_{\vect{R}}^{\alpha} v^{\alpha} (\vect{x} - \vect{R}^{\alpha},
   \vect{x}' - \vect{R}^{\alpha}) ,
\end{equation}
in which $\vect{R}$ is now a Bravais lattice vector (for the periodic
bulk reference crystal), $\alpha$ labels an atom at position
$\btau^{\alpha}$ in the unit cell, and $\vect{R}^{\alpha} = \vect{R} +
\btau^{\alpha}$.  The function $f_{\vect{R}}^{\alpha}$ is the atomic
mole fraction of atom $\alpha$ at position $\vect{R}^{\alpha}$, while
$v^{\alpha} (\vect{x}, \vect{x}')$ is the nonlocal ionic
pseudopotential for atom $\alpha$.

With the coupling to the magnetic field given by Eqs.\ (\ref{eq:VA})
and (\ref{eq:path2}), the Fourier transform of the field-dependent
potential can be written as
\begin{equation}
   \tilde{V} (\vect{k}, \vect{k}') = \sum_{\alpha, \vect{R}}
   f_{\vect{R}}^{\alpha} \rme^{-\rmi (\vect{k} - \vect{k}') \cdot
   \vect{R}^{\alpha}} \exp [\rmi \xi (\hat{\vect{x}}, \hat{\vect{x}}')
   / c] v^{\alpha} (\vect{K}, \vect{K}') , \label{eq:Vkk}
\end{equation}
in which $\vect{K}$ is the kinetic momentum
\begin{equation}
   \vect{K} = \vect{k} + \frac{1}{c}\vect{A} (\vect{R}^{\alpha}) ,
   \qquad \vect{K}' = \vect{k}' + \frac{1}{c}\vect{A}
   (\vect{R}^{\alpha}) .
\end{equation}
The operator function $\xi$ is defined by
\begin{equation}
   \fl \xi (\hat{\vect{x}}, \hat{\vect{x}}') = \frac{1}{2!}
   (\hat{x}_{\lambda}' \hat{x}_{\mu}' - \hat{x}_{\lambda}
   \hat{x}_{\mu}) \frac{\partial A_{\lambda}}{\partial
   R^{\alpha}_{\mu}} + \frac{1}{3!}  (\hat{x}_{\lambda}'
   \hat{x}_{\mu}' \hat{x}_{\nu}' - \hat{x}_{\lambda} \hat{x}_{\mu}
   \hat{x}_{\nu}) \frac{\partial^2 A_{\lambda}}{\partial
   R^{\alpha}_{\mu} \partial R^{\alpha}_{\nu}} + \cdots ,
   \label{eq:xi}
\end{equation}
in which $\hat{x}_{\lambda} = \rmi \partial / \partial k_{\lambda}$
and $\hat{x}_{\mu}' = -\rmi \partial / \partial k_{\mu}'$.  These
expressions are used to develop an effective-mass Hamiltonian in what
follows.

\section{Approximations to the Hamiltonian}

\label{sec:approx}

Effective-mass theory is based on approximations valid for
$k_{\mathrm{env}} a \ll 1$, where $k_{\mathrm{env}}$ is a typical wave
number for a slowly varying envelope function.  This condition will
hold, for example, in a wide quantum well of width $L \gg a$, for
which $k_{\mathrm{env}} \sim 2 \pi / L$.  The external magnetic field
$B$ must also be weak enough for the magnetic length $L_B =
\sqrt{c/B}$ to satisfy $L_B \gg a$ \cite{BirPik74}.

In addition to these standard conditions, if $\vect{K}$ is to be
treated as a small quantity of the same order as $\vect{k}$, the
vector potential must be small enough to satisfy $a A / c \ll 1$.  For
a sinusoidal field with $B \sim k_B A$ \cite{MauLou96}, where $k_B$ is
the wave number of the external field, this condition is equivalent to
$k_B L_B^2 \gg a$.  Combining this with the previous condition $L_B
\gg a$, the new condition will be satisfied if $k_B L_B
\raisebox{-0.9ex}{ $\overset{\textstyle >}{\sim}$ } 1$.  Thus, very
small values of $k_B$ place an additional constraint on $L_B$.  This
constraint is not usually mentioned \cite{BirPik74}, but it is
implicit in any effective-mass treatment where $\vect{K}$ is assumed
to be of the same order as $\vect{k}$.  Therefore, effective-mass
calculations for truly constant $\vect{B}$ fields (which are obtained
in the limit $k_B \rightarrow 0$ \cite{MauLou96}) can only yield
results that are valid asymptotically in the limit $B \rightarrow 0$.
The asymptotic nature of effective-mass theory for constant $\vect{B}$
was suggested by Kohn \cite{Kohn59b}, and stated more explicitly by
Foldy and Wouthuysen \cite{FolWou50} in connection with the
nonrelativistic approximation to the Dirac equation.

In the Luttinger-Kohn representation $|n \vect{k} \rangle$
\cite{LuKo55}, the wave function $\langle n \vect{k} | \psi \rangle$
is just a unitary transformation of the momentum-space wave function
$\langle \vect{k} + \vect{G} | \psi \rangle$ \cite{Burt92}:
\begin{equation}
   \langle n \vect{k} | \psi \rangle = \sum_{\vect{G}}
   U_{n\vect{G}}^{\dag} \langle \vect{k} + \vect{G} | \psi \rangle .
\end{equation}
Here $\vect{G}$ is a reciprocal lattice vector of the reference
crystal, and the spinor $U_{n\vect{G}}$ is a Fourier series
coefficient for the $n^{\mathrm{th}}$ zone-center Bloch function of
the reference crystal.  For slowly varying envelope functions $F_n
(\vect{x})$, the Fourier transform $F_n (\vect{k}) \equiv \langle n
\vect{k} | \psi \rangle$ (and hence $\langle \vect{k} + \vect{G} |
\psi \rangle$) is significant only for small $k$.  (A review of the
numerical evidence supporting the use of the slowly varying envelope
approximation in abrupt heterostructures is given in reference
\cite{Fore05b}.)  Thus, slowly varying envelope functions probe the
nonlocal pseudopotential only within a small neighborhood of the
reciprocal lattice vectors $\vect{G}$.  Within this neighborhood, the
ionic pseudopotential can be approximated by the leading terms in its
Taylor series expansion \cite{Fore05b}:
\begin{eqnarray}
   \fl v^{\alpha} (\vect{k} + \vect{G}, \vect{k}' + \vect{G}') =
   v^{(\cdot| \cdot) \alpha}_{\vect{G} \vect{G}'} + k_{\lambda}
   v^{(\lambda|\cdot) \alpha}_{\vect{G} \vect{G}'} + k_{\lambda}'
   v^{(\cdot|\lambda) \alpha}_{\vect{G} \vect{G}'} + k_{\lambda}
   k_{\mu} v^{(\lambda\mu|\cdot) \alpha}_{\vect{G} \vect{G}'} +
   k_{\lambda}' k_{\mu}' v^{(\cdot|\lambda\mu) \alpha}_{\vect{G}
   \vect{G}'} \nonumber \\ {} + k_{\lambda} k_{\mu}' v^{(\lambda|\mu)
   \alpha}_{\vect{G} \vect{G}'} + \cdots , \label{eq:vTaylor}
\end{eqnarray}
where $v^{(\lambda|\mu) \alpha}_{\vect{G} \vect{G}'}$ is an expansion
coefficient.  This expansion is valid because $v^{\alpha} (\vect{x},
\vect{x}')$ is localized, hence $v^{\alpha} (\vect{k}, \vect{k}')$ is
analytic.  For simplicity, terms beyond the second order are not
written out explicitly here; however, the present effective-mass
theory \cite{Fore05b} retains terms of order $k^4$ in the Hamiltonian
of the reference crystal [see equation (\ref{eq:Vxp}) below].

By definition, the expansion coefficients are symmetric with respect
to arbitrary permutations of the indices on one side of the vertical
line---for example, $v^{(\lambda\mu|\cdot) \alpha}_{\vect{G}
\vect{G}'} = v^{(\mu\lambda|\cdot) \alpha}_{\vect{G} \vect{G}'}$.
However, $v^{(\lambda|\mu) \alpha}_{\vect{G} \vect{G}'} \ne
v^{(\mu|\lambda) \alpha}_{\vect{G} \vect{G}'}$.  For indices on
opposite sides of the vertical line, the only symmetries are those
obtained from hermiticity and time reversal, such as $v^{(\lambda|\mu)
\alpha}_{\vect{G} \vect{G}'} = (v^{(\mu|\lambda) \alpha}_{\vect{G}'
\vect{G}})^{\dag}$ and $v^{(\cdot|\lambda) \alpha}_{\vect{G}
\vect{G}'} = (v^{(\lambda|\cdot) \alpha}_{\vect{G}'
\vect{G}})^{\dag}$.

With the above restrictions on the magnetic field, $\vect{K}$ is a
small quantity of the same order as $\vect{k}$, and the potential
(\ref{eq:Vkk}) can be expanded in a Taylor series about the reciprocal
lattice vectors $\vect{G}$.  This yields
\begin{eqnarray}
   \fl \tilde{V} (\vect{k} + \vect{G}, \vect{k}' + \vect{G}') =
   \sum_{\alpha, \vect{R}} f_{\vect{R}}^{\alpha} \rme^{-\rmi (\vect{k}
   - \vect{k}') \cdot \vect{R}^{\alpha}} e^{-\rmi (\vect{G} -
   \vect{G}') \cdot \btau^{\alpha}} \Biggl[ v^{(\cdot| \cdot)
   \alpha}_{\vect{G} \vect{G}'} + K_{\lambda} v^{(\lambda|\cdot)
   \alpha}_{\vect{G} \vect{G}'} + K_{\lambda}' v^{(\cdot|\lambda)
   \alpha}_{\vect{G} \vect{G}'} \nonumber \\ {} + \Biggl( K_{\lambda}
   K_{\mu} + \frac{\rmi}{c} \frac{\partial A_{\lambda}}{\partial
   R^{\alpha}_{\mu}} \Biggr) v^{(\lambda\mu|\cdot) \alpha}_{\vect{G}
   \vect{G}'} + \Biggl( K_{\lambda}' K_{\mu}' - \frac{\rmi}{c}
   \frac{\partial A_{\lambda}}{\partial R^{\alpha}_{\mu}} \Biggr)
   v^{(\cdot|\lambda\mu) \alpha}_{\vect{G} \vect{G}'} \nonumber \\ {}
   + K_{\lambda} K_{\mu}' v^{(\lambda|\mu) \alpha}_{\vect{G}
   \vect{G}'} + \cdots \Biggr] , \label{eq:VkG}
\end{eqnarray}
where the expansion coefficients are the same as in
(\ref{eq:vTaylor}), and the terms involving $\partial A_{\lambda} /
\partial R^{\alpha}_{\mu}$ are the leading contributions from
(\ref{eq:xi}).

\section{Envelope-function Hamiltonian}

\label{sec:ef}

These results are now used to construct an envelope-function
Hamiltonian valid for slowly varying envelopes.  In the Luttinger-Kohn
representation, the matrix elements of $\tilde{V}$ are given by the
unitary transformation
\begin{equation}
   \tilde{V}_{nn'} (\vect{k}, \vect{k}') = \sum_{\vect{G},\vect{G}'}
   U_{n\vect{G}}^{\dag} \tilde{V} (\vect{k} + \vect{G}, \vect{k}' +
   \vect{G}') U_{n'\vect{G}'} . \label{eq:Vnnkk}
\end{equation}
When the expansion (\ref{eq:VkG}) is inserted into equation
(\ref{eq:Vnnkk}), one obtains transformed expansion coefficients such
as
\begin{equation}
   V^{(\lambda\mu|\cdot)\alpha}_{nn'} = N \sum_{\vect{G},\vect{G}'}
   U_{n\vect{G}}^{\dag}
   v^{(\lambda\mu|\cdot)\alpha}_{\vect{G}\vect{G}'} U_{n'\vect{G}'}
   e^{-\rmi (\vect{G} - \vect{G}') \cdot \btau^{\alpha}} ,
\end{equation}
in which $N$ is the number of unit cells in the crystal.  It is also
convenient to introduce coordinate-space matrix elements of the form
\begin{equation}
   V^{(\lambda\mu|\cdot)}_{nn'} (\vect{x}) = \sum_{\alpha}
   V^{(\lambda\mu|\cdot)\alpha}_{nn'} f^{\alpha} (\vect{x}) ,
\end{equation}
in which $f^{\alpha} (\vect{x})$ is a macroscopic average (see
reference \cite{Fore05b}) of the discrete atomic distribution function
$f^{\alpha}_{\vect{R}}$.  Note that $V^{(\lambda\mu|\cdot)}_{nn'}
(\vect{x}) = V^{(\mu\lambda|\cdot)}_{nn'} (\vect{x}) =
[V^{(\cdot|\lambda\mu)}_{n'n} (\vect{x})]^*$ and
$V^{(\lambda|\mu)}_{nn'} (\vect{x}) = [V^{(\mu|\lambda)}_{n'n}
(\vect{x})]^*$.

With these notational conventions, the Fourier transform of the matrix
element (\ref{eq:Vnnkk}) yields the following contribution from the
nonlocal pseudopotential to the envelope-function Hamiltonian:
\begin{eqnarray}
   \fl \tilde{V}_{nn'}(\vect{x}, \vect{p}) = V^{(\cdot|\cdot)}_{nn'}
   (\vect{x}) + P_{\lambda} V^{(\lambda|\cdot)}_{nn'} (\vect{x}) +
   V^{(\cdot|\lambda)}_{nn'} (\vect{x}) P_{\lambda} + P_{\lambda}
   P_{\mu} V_{nn'}^{(\lambda\mu|\cdot)} (\vect{x}) +
   V_{nn'}^{(\cdot|\lambda\mu)} (\vect{x}) P_{\lambda} P_{\mu}
   \nonumber \\ {} + P_{\lambda} V_{nn'}^{(\lambda|\mu)} (\vect{x})
   P_{\mu} + V_{nn'}^{\lambda\mu\nu} P_{\lambda} P_{\mu} P_{\nu} +
   V_{nn'}^{\lambda\mu\kappa\nu} P_{\lambda} P_{\mu} P_{\kappa}
   P_{\nu} , \label{eq:Vxp}
\end{eqnarray}
in which the canonical and kinetic momentum operators are $\vect{p} =
-\rmi \nabla$ and $\vect{P} = \vect{p} + \vect{A} (\vect{x}) / c$.
The derivation of this result makes use of the usual approximations
\cite{BirPik74} for slowly varying potentials $\vect{A} (\vect{x})$
and envelope functions $F_n (\vect{x})$.  In other words, it was
assumed that since only small values of $\vect{k}$ are of interest,
the error involved in replacing the Fourier transform of the discrete
product $f^{\alpha}_{\vect{R}} \vect{A} (\vect{R}_{\alpha})$ with the
Fourier transform of the continuous product $f^{\alpha} (\vect{x})
\vect{A} (\vect{x})$ is negligible, and the upper limit of any
convolutions in $\vect{k}$ space can be extended to infinity.  This
may be called the {\em local} approximation because it reduces
equation (\ref{eq:Vxp}) to the form of a local differential operator;
see reference \cite{Fore05b} for a discussion of the error involved in
this approximation.

In keeping with the perturbation theory of reference \cite{Fore05b},
the full position dependence of $f^{\alpha}_{\vect{R}}$ was retained
in the terms quadratic in $\vect{P}$ in equation (\ref{eq:Vxp}).
However, in the cubic and quartic terms, $f^{\alpha}_{\vect{R}}$ was
replaced by the atomic distribution function $f^{\alpha}$ of the
reference crystal (i.e., the heterostructure perturbation
$\theta^{\alpha}_{\vect{R}} = f^{\alpha}_{\vect{R}} - f^{\alpha}$ was
neglected).  Thus, the coefficients of the cubic and quartic terms are
given by the constants
\begin{eqnarray}
   V_{nn'}^{\lambda\mu\nu} = V_{nn'}^{(\lambda\mu\nu|\cdot)} +
   V_{nn'}^{(\lambda\mu|\nu)} + V_{nn'}^{(\lambda|\mu\nu)} +
   V_{nn'}^{(\cdot|\lambda\mu\nu)} , \label{eq:Vcubic} \\
   V_{nn'}^{\lambda\mu\kappa\nu} =
   V_{nn'}^{(\lambda\mu\kappa\nu|\cdot)} +
   V_{nn'}^{(\lambda\mu\kappa|\nu)} + V_{nn'}^{(\lambda\mu|\kappa\nu)}
   + V_{nn'}^{(\lambda|\mu\kappa\nu)} +
   V_{nn'}^{(\cdot|\lambda\mu\kappa\nu)} , \label{eq:Vquartic}
\end{eqnarray}
in which the various terms are just expansion coefficients that were
omitted in equation (\ref{eq:vTaylor}).

Because of the general lack of symmetry of the expansion coefficients
in equation (\ref{eq:vTaylor}), the noncommuting operators in equation
(\ref{eq:Vxp}) have a definite operator ordering.  The term involving
$V_{nn'}^{(\lambda|\mu)} (\vect{x})$ has the same operator ordering as
the Hamiltonian studied in the work of Mlinar \etal \cite{Mlinar05},
and includes terms antisymmetric in $\lambda$ and $\mu$.  However,
that Hamiltonian \cite{Fore93,Fore97} does not include any terms with
the operator ordering of $V_{nn'}^{(\lambda\mu|\cdot)} (\vect{x})$ and
$V_{nn'}^{(\cdot|\lambda\mu)} (\vect{x})$.

As might be expected from the work of Luttinger \cite{Lutt56},
equation (\ref{eq:Vxp}) is not the same as what would be obtained by
substituting $\vect{p} \rightarrow \vect{P}$ into the Hamiltonian of
reference \cite{Fore05b}.  The position-dependent terms describing the
linear response to the heterostructure perturbation do have this form,
since care was taken in reference \cite{Fore05b} to retain the
asymmetric parts of these material parameters.  However, the bulk
coefficients derived here [such as (\ref{eq:Vcubic}),
(\ref{eq:Vquartic}), and the bulk part of $V^{(\lambda|\mu)}_{nn'}$]
are asymmetric in the Cartesian indices, whereas those obtained from
the expansion of $v^{\alpha}_{\vect{G}\vect{G}'} (\vect{k}) \equiv
v^{\alpha} (\vect{k} + \vect{G}, \vect{k} + \vect{G}')$ in reference
\cite{Fore05b} were symmetric.  These asymmetric coefficients generate
terms in the Hamiltonian proportional to the magnetic field, such as
the dipole term in equation (\ref{eq:emxP}) below.

\section{Other contributions to the envelope-function Hamiltonian}

\label{sec:other}

Within LDA, the self-energy of the electron is a local potential that
does not couple to a static magnetic field \cite{PiMa03}.  Thus, the
only other term in the Hamiltonian that depends on the magnetic field
is the kinetic energy $\frac12 \vect{p}^2$.  To account for the
intrinsic dipole moment of the electron, this can be written as
$\frac12 (\bsigma \cdot \vect{p})^2$, where $\bsigma$ is the Pauli
matrix vector; the minimal substitution $\vect{p} \rightarrow
\vect{P}$ then generates the electron $g$ factor $g_0 = 2$
automatically \cite{Sak67_p78}.

In the Luttinger-Kohn representation, the kinetic energy produces
$\vect{k} \cdot \vect{p}$ terms of the usual form
\cite{LuKo55,Lutt56,BirPik74,Leib77}; a finite-dimensional multi-band
effective-mass equation can then be derived using the standard
canonical transformation method of perturbation theory
\cite{Wink03,LuKo55,BirPik74,Leib77}, with the $\vect{k} \cdot
\vect{p}$ interaction and the heterostructure potential treated as
perturbations \cite{Fore05b,Leib77}.  The details are essentially the
same as those given in reference \cite{Fore05b}, which will not be
repeated here.  The outcome is a Hamiltonian with the same qualitative
structure as that shown in equation (\ref{eq:Vxp}), but with all of
the coefficients renormalized [e.g., $V^{(\lambda\mu|\cdot)}_{nn'}
(\vect{x}) \rightarrow \bar{V}^{(\lambda\mu|\cdot)}_{nn'} (\vect{x})$]
to account for perturbative corrections.

This Hamiltonian can then be rearranged into a more convenient form
\cite{Leib77}, which yields the effective-mass Hamiltonian
\begin{eqnarray}
   \fl \tilde{H}_{nn'} (\vect{x}, \vect{p}) = E_n \delta_{nn'} + \{
   P_{\lambda} , \pi^{\lambda}_{nn'} (\vect{x}) \} + \{ \{ P_{\lambda}
   P_{\mu} \} , D^{ \{ \lambda\mu \} }_{nn'} (\vect{x}) \} -
   \mu^{\lambda}_{nn'} (\vect{x}) B_{\lambda} (\vect{x}) \nonumber \\
   {} + P_{\lambda} P_{\mu} P_{\kappa} C^{\lambda\mu\kappa}_{nn'} +
   P_{\lambda} P_{\mu} P_{\kappa} P_{\nu}
   Q^{\lambda\mu\kappa\nu}_{nn'} + W_{nn'} (\vect{x}) + \varphi
   (\vect{x}) \delta_{nn'} \nonumber \\ {} + \partial_{\lambda}
   Z^{\lambda}_{nn'} (\vect{x}) + \partial_{\lambda} \partial_{\mu}
   Y^{\lambda\mu}_{nn'} (\vect{x}) + \{ P_{\lambda} , \partial_{\mu}
   \Gamma^{\lambda\mu}_{nn'} (\vect{x}) \} + [ \partial_{\mu}
   \Phi^{\lambda\mu}_{nn'} (\vect{x}) ] P_{\lambda} \nonumber \\ {} +
   \hat{Z}^{\lambda}_{nn'} \partial_{\lambda} \varphi (\vect{x}) +
   \hat{Y}^{\lambda\mu}_{nn'} \partial_{\lambda} \partial_{\mu}
   \varphi (\vect{x}) + \hat{\Gamma}^{\lambda\mu}_{nn'} \{ P_{\lambda}
   , \partial_{\mu} \varphi (\vect{x}) \} +
   \hat{\Phi}^{\lambda\mu}_{nn'} [ \partial_{\mu} \varphi (\vect{x}) ]
   P_{\lambda} , \label{eq:emxP}
\end{eqnarray}
in which $\{ A, B \} = \{ A B \} = \frac12 (AB + BA)$ is the
symmetrized product, and $\partial_{\lambda} = \partial / \partial
x_{\lambda}$ acts only on the function immediately to its right.  The
various coefficients in this equation are the same as those defined in
reference \cite{Fore05b}, with some modifications to be discussed
below.  Therefore, the definitions given in \cite{Fore05b} are not
repeated here.

The first term $E_n$ is just the zone-center energy of the reference
crystal, while $\pi^{\lambda}_{nn'} (\vect{x})$ is the
position-dependent momentum matrix of the heterostructure.  The
Luttinger $D$ matrix \cite{Lutt56}, which is half the inverse
effective mass tensor, is defined by
\begin{equation}
   \fl D_{nn'}^{\lambda\mu} (\vect{x}) = \frac12 (\delta_{\lambda\mu}
   \delta_{nn'} + \rmi \epsilon_{\lambda\mu\nu} \sigma^{\nu}_{nn'}) +
   \bar{V}^{(\lambda\mu|\cdot)}_{nn'} (\vect{x}) +
   \bar{V}^{(\cdot|\lambda\mu)}_{nn'} (\vect{x}) +
   \bar{V}^{(\lambda|\mu)}_{nn'} (\vect{x}) , \label{eq:Dnn1}
\end{equation}
in which $\epsilon_{\lambda\mu\nu}$ is the antisymmetric unit tensor.
Functions such as $\bar{V}^{(\lambda\mu|\cdot)}_{nn'} (\vect{x})$ have
the same form as $V^{(\lambda\mu|\cdot)}_{nn'} (\vect{x})$, but
include additional $\vect{k} \cdot \vect{p}$ renormalization terms
that are given in \cite{Fore05b}.  However,
$\bar{V}^{(\lambda\mu|\cdot)}_{nn'} (\vect{x})$ is no longer
symmetric: $\bar{V}^{(\lambda\mu|\cdot)}_{nn'} (\vect{x}) \ne
\bar{V}^{(\mu\lambda|\cdot)}_{nn'} (\vect{x})$.

The symmetric part of the tensor (\ref{eq:Dnn1}) is written as
$D_{nn'}^{\{\lambda\mu\}} (\vect{x})$, while the antisymmetric part
determines the effective magnetic dipole moment
\begin{equation}
   \mu^{\lambda}_{nn'} (\vect{x}) = \frac{\rmi}{2c}
   \epsilon_{\lambda\mu\nu} D^{\mu\nu}_{nn'} (\vect{x}) .
   \label{eq:dipole}
\end{equation}
The symmetric terms $V^{(\lambda\mu|\cdot)}_{nn'} (\vect{x})$ and
$V^{(\cdot|\lambda\mu)}_{nn'} (\vect{x})$ do not contribute to
(\ref{eq:dipole}); the sole contribution from the nonlocal potential
is the asymmetric term $V^{(\lambda|\mu)}_{nn'} (\vect{x})$.  Note,
however, that this result depends on the initial choice of
(\ref{eq:VA}) as the magnetic coupling Hamiltonian.  As shown in the
Appendix, for the generalized Peierls coupling of Ismail-Beigi \etal
\cite{IsChLouie01}, the nonlocal potential contributes nothing to the
dipole moment (\ref{eq:dipole}).

The cubic ($C^{\lambda\mu\kappa}_{nn'}$) and quartic
($Q^{\lambda\mu\kappa\nu}_{nn'}$) dispersion terms are just
renormalized versions of (\ref{eq:Vcubic}) and (\ref{eq:Vquartic}).
The term $W_{nn'} (\vect{x})$ represents linear and quadratic
contributions to the effective potential energy, one portion of which
is $V^{(\cdot|\cdot)}_{nn'} (\vect{x})$ from equation (\ref{eq:Vxp}).
The screened potential $\varphi (\vect{x})$ accounts for long-range
multipole potentials generated by the electron-electron Coulomb
interaction \cite{Fore05b}.

The terms $Z^{\lambda}_{nn'} (\vect{x})$, $Y^{\lambda\mu}_{nn'}
(\vect{x})$, $\Gamma^{\lambda\mu}_{nn'} (\vect{x})$, and
$\Phi^{\lambda\mu}_{nn'} (\vect{x})$ are short-range interface terms,
the contribution from which vanishes in bulk material because the
spatial derivatives $\partial_{\lambda}$ are zero there.  $Z$
generates a $\delta$-like interface mixing of light and heavy holes
\cite{IvKaRo96}, while $Y$ is similar to the Darwin term in the
nonrelativistic approximation to the Dirac equation.  The term $\Phi$,
which is antisymmetric, is a generalized Rashba coupling \cite{Wink03}
that can be viewed as an effective spin-orbit interaction.  This is
precisely the antisymmetric term that was examined in the work of
Mlinar \etal \cite{Mlinar05}.  The $\Gamma$ term has a similar
structure (i.e., an interface term that is linear in $\vect{P}$) but
is symmetric.

The contributions from $\hat{Z}$, $\hat{Y}$, $\hat{\Gamma}$, and
$\hat{\Phi}$ are similar to those from $Z$, $Y$, $\Gamma$, and $\Phi$,
but since they involve the long-range multipole potential $\varphi
(\vect{x})$, they are not as well localized at the interface
\cite{Fore05b}.  These terms are not included in most
envelope-function theories.  However, in a quantum well or other
two-dimensional system, these terms have no qualitative effect, as
they merely renormalize the short-range contributions \cite{Fore05b}.
This is not strictly true in quantum wires or dots, but even in these
cases it may be possible to replace the long-range contributions with
renormalized short-range terms, since Zunger \etal
\cite{MaZu94,WaZu95,FuZu97} have shown that LDA wave functions in a
variety of nanostructures can be accurately reproduced using carefully
fitted empirical pseudopotentials, which by definition have no
long-range multipole terms.

\section{Conclusions and discussion}

\label{sec:conclusions}

In this paper, a first-principles multi-band effective-mass
Hamiltonian was derived for lattice-matched semiconductor
nanostructures in a slowly varying magnetic field.  The theory applies
to systems described by norm-conserving nonlocal pseudopotentials in
the local density approximation, and is based on the magnetic coupling
Hamiltonian derived by Pickard and Mauri \cite{PiMa03}.  The main
result, given in equation (\ref{eq:emxP}), is similar to what would be
obtained by applying the minimal substitution $\vect{p} \rightarrow
\vect{p} + \vect{A} / c$ to the Hamiltonian derived in reference
\cite{Fore05b}.  However, there were also several additional terms,
derived from asymmetric matrix elements of the nonlocal
pseudopotential, that were not included in reference \cite{Fore05b}.
As shown in the Appendix, these terms are controlled by the choice of
magnetic coupling for the nonlocal pseudopotential.  The present work
provides the first direct analytical demonstration of the difference
between the magnetic coupling Hamiltonians of Pickard and Mauri
\cite{PiMa03} and Ismail-Beigi \etal \cite{IsChLouie01}.  Namely, in
the former case \cite{PiMa03}, the nonlocal potential contributes to
the effective magnetic dipole moment, while in the latter case
\cite{IsChLouie01} it does not.

Turning now to the significance of the various interface terms in the
Hamiltonian (\ref{eq:emxP}), the work of Mlinar \etal \cite{Mlinar05}
has established the importance of including the generalized Rashba
coupling $\Phi^{\lambda\mu}_{nn'} (\vect{x})$.  In their calculations,
the magnitude of this term was estimated using the expressions given
in references \cite{Fore93} and \cite{Fore97}, which are equivalent to
the assumption that $\Phi^{\lambda\mu}_{nn'} (\vect{x})$ is the same
as the antisymmetric part $\rmi D^{[\lambda\mu]}_{nn'} (\vect{x})$ of
the effective mass tensor (\ref{eq:Dnn1}), or that
$\Phi^{\sigma\tau}_{nn'} (\vect{x}) = c \epsilon_{\sigma\tau\nu}
\mu^{\nu}_{nn'} (\vect{x})$.  (Note that for $\Gamma_8$ states in
zinc-blende, $\mu^{\nu} (\vect{x}) = [\kappa (\vect{x}) J_{\nu} + q
(\vect{x}) J_{\nu}^3] / c$, where $\vect{J}$ is an angular momentum
$\frac32$ matrix, and $\kappa$ and $q$ are Luttinger parameters
\cite{Lutt56}.  The calculations in reference \cite{Mlinar05} included
only the contribution from $\kappa$.)  As discussed in reference
\cite{Fore05b}, this assumption provides a reasonable
order-of-magnitude estimate, but cannot be relied upon for
quantitative accuracy.  Thus, the large qualitative impact of
$\Phi^{\lambda\mu}_{nn'} (\vect{x})$ discovered by Mlinar \etal
\cite{Mlinar05} is almost certainly a real physical effect, even
though the quantitative details of their calculation may not be
completely accurate.

In addition to $\Phi^{\lambda\mu}_{nn'} (\vect{x})$, the Hamiltonian
(\ref{eq:emxP}) also includes the interface terms $Z^{\lambda}_{nn'}
(\vect{x})$, $Y^{\lambda\mu}_{nn'} (\vect{x})$, and
$\Gamma^{\lambda\mu}_{nn'} (\vect{x})$, which were not studied by
Mlinar \etal \cite{Mlinar05}.  A quantitative calculation of these
matrix elements is now in progress (and will be reported elsewhere),
but a rough idea of their significance can be obtained from
dimensional and symmetry analysis.  The terms $Z^{\lambda}_{nn'}
(\vect{x})$ and $Y^{\lambda\mu}_{nn'} (\vect{x})$ affect the boundary
conditions on the envelope functions, but do not contain any direct
contribution from the magnetic field.  Therefore, their influence on
the magnetic-field-dependent properties of nanostructures is probably
not as strong as that of $\Phi^{\lambda\mu}_{nn'} (\vect{x})$.
Nevertheless, their contribution may still be significant---especially
the $\delta$-function term $Z^{\lambda}_{nn'} (\vect{x})$
\cite{IvKaRo96}, which is of lower order \cite{Fore05b} than the other
interface terms.

On the other hand, the term involving $\Gamma^{\lambda\mu}_{nn'}
(\vect{x})$, like $\Phi^{\lambda\mu}_{nn'} (\vect{x})$, is linear in
the vector potential.  The symmetry analysis of reference
\cite{Fore05b} shows that this term is not present in the $\Gamma_6$
conduction band of zinc-blende materials, and that in the $\Gamma_8$
valence band it is a relativistic effect that vanishes when spin-orbit
coupling is neglected.  Its magnitude will therefore be smaller than
that of $\Phi^{\lambda\mu}_{nn'} (\vect{x})$, but it may be important
in the description of spin-splitting effects.

Therefore, even though the calculations of Mlinar \etal
\cite{Mlinar05} have included what is probably the dominant
magnetic-field-dependent interface term, the remaining terms are not
likely to be negligible.  The detailed study of these terms, however,
is left to future numerical work.

\ack

This work was supported by Hong Kong UGC grant number HIA03/04.SC02.

\appendix

\section*{Appendix}

\setcounter{section}{1}

\label{app:ICL}

This appendix examines the consequences of replacing the coupling
Hamiltonian of Pickard and Mauri \cite{PiMa03} with that of
Ismail-Beigi \etal \cite{IsChLouie01}, who have proposed a
generalization of the well known Peierls phase \cite{Pei33,Pei97}:
\begin{equation}
   \tilde{V}^{{\rm nl}} (\vect{x}, \vect{x}') = V^{{\rm nl}}
   (\vect{x}, \vect{x}') \exp \Biggl( \frac{\rmi}{c} \int_{\vect{x}
   \rightarrow \vect{x}'} \vect{A} ( \vect{x}'' ) \cdot \rmd
   \vect{x}'' \Biggr) .  \label{eq:VICL}
\end{equation}
Unlike equation (\ref{eq:VA}), this does not rely on separating
$V^{{\rm nl}} (\vect{x}, \vect{x}')$ into a sum of localized atomic
potentials---although, for the application considered here, it is
assumed that the potential has this form.  The line integral can be
evaluated by replacing $\vect{R}$ with $\vect{x}$ in equation
(\ref{eq:path1}) and then expanding $\vect{A} (\vect{x})$ in a Taylor
series about $\vect{x} = \vect{R}$, which gives
\begin{eqnarray}
   \fl \int_{\vect{x} \rightarrow \vect{x}'} \vect{A} \cdot \rmd
   \vect{x}'' = (\vect{x}' - \vect{x}) \cdot \vect{A} (\vect{R}) +
   \frac{1}{2} [ (x_{\lambda}' - R_{\lambda}) (x_{\mu}' - R_{\mu}) -
   (x_{\lambda} - R_{\lambda}) (x_{\mu} - R_{\mu}) ] \frac{\partial
   A_{\lambda}}{\partial R_{\mu}} \nonumber \\ {} - \frac12
   (x_{\lambda} - R_{\lambda}) (x_{\mu}' - R_{\mu}) \Biggl(
   \frac{\partial A_{\lambda}}{\partial R_{\mu}} - \frac{\partial
   A_{\mu}}{\partial R_{\lambda}} \Biggr) + \cdots .  \label{eq:path3}
\end{eqnarray}
The last term, which was not present in equation (\ref{eq:path2}), is
proportional to the magnetic field.

If the derivation in sections \ref{sec:approx} and \ref{sec:ef} is now
carried through in the same way as before, one finds that the last
term shown in equation (\ref{eq:VkG}) is modified as follows:
\begin{equation}
   K_{\lambda} K_{\mu}' v^{(\lambda|\mu) \alpha}_{\vect{G} \vect{G}'}
   \rightarrow \Biggl( K_{\lambda} K_{\mu}' + \frac{\rmi}{2c}
   \epsilon_{\lambda\mu\nu} B_{\nu} (\vect{R}^{\alpha}) \Biggr)
   v^{(\lambda|\mu) \alpha}_{\vect{G} \vect{G}'} .
\end{equation}
The effect of the additional term is to cancel the contribution from
$V^{(\lambda|\mu)}_{nn'} (\vect{x})$ to the effective dipole moment
(\ref{eq:dipole}).  Thus, for this coupling, the nonlocal potential
makes no contribution to the dipole moment.  A similar conclusion was
reached by Kane \cite{Kane58}, who used the Peierls coupling
(\ref{eq:VICL}) for the special case of a constant magnetic field.

Therefore, different choices of coupling between the nonlocal
potential and the magnetic field have a direct impact on the effective
dipole moment of the electron.  This provides an explicit analytical
demonstration of the differences demonstrated numerically in reference
\cite{PiMa03}.

\section*{References}


\end{document}